\documentclass[12pt]{article}
\usepackage[utf8]{inputenc}
\usepackage{mathptmx}
\usepackage{geometry}
\usepackage{authblk}
\usepackage{amssymb}
\usepackage{amsmath}
\usepackage{amsthm}
\allowdisplaybreaks[4]
\usepackage{slashed}
\usepackage{graphicx,color,epsfig}
\usepackage{multirow}
\usepackage{cite}
\usepackage[colorlinks=true,linkcolor=red,citecolor=blue]{hyperref}
\begin{document}
\renewcommand{\arraystretch}{1.2}
\newcommand{\beq}{\begin{eqnarray}}
\newcommand{\eeq}{\end{eqnarray}}
\newcommand{\non}{\nonumber\\ }
\newcommand{\acp}{ {\cal A}_{CP} }
\newcommand{\psl}{ p \hspace{-1.8truemm}/ }
\newcommand{\nsl}{ n \hspace{-2.2truemm}/ }
\newcommand{\vsl}{ v \hspace{-2.2truemm}/ }
\newcommand{\epsl}{\epsilon \hspace{-1.8truemm}/\,  }
\def\ra{\rangle}
\def\la{\langle}
\def\sl{\!\!\!\!\slash}
\def\ov{\overline}
\newcommand{\tf}{\textbf}
\title{Family Non-universal $Z^\prime$ Effects on $B_{d,s} \to K^{*0} {\overline K^{*0}}$ Decays in Perturbative QCD Approach}

\author[1]{Ying Li \footnote{liying@ytu.edu.cn} }
\author[2]{Guo-Hua Zhao}
\author[2]{Yan-Jun Sun  \footnote{sunyanjun@nwnu.edu.cn}}
\author[1]{Zhi-Tian Zou}
\affil[1]{\it \small Department of Physics, Yantai University, Yantai 264005, China}
\affil[2]{\it \small Department of Physics, North-west Normal University, Lanzhou 221116,China}
\maketitle
\begin{abstract}
The nonleptonic decays $B_{d,s} \to K^{*0} {\overline K^{*0}}$ are reanalyzed in perturbative QCD approach, which is based on the $k_{\rm T}$ factorization. In the standard model, the calculated branching fraction and longitudinal polarization fraction of $B_{d} \to K^{*0} {\overline K^{*0}}$ are in agreement with experimental measurements, while the predictions of $B_{s} \to K^{*0} {\overline K^{*0}}$ cannot agree with data simultaneously. The parameter that combines of longitudinal polarization fractions and branching fractions evaluated to be $L_{K^*\overline{K}^{*0}}^{\rm PQCD}= 12.7^{+5.6} _{-3.2}$, which is also larger than that abstracted from experimental measurements. We then study all observables by introducing a family non-universal $Z^{\prime}$ boson in $b\to s q\bar q$ transitions. In order to reduce the number of new parameters, we simplify the model as possible. It is found that with the fixed value $\omega_{B_s}=0.55$, these exists parameter space where all measurements, including the branching fraction, longitudinal polarization fraction and $L_{K^*\overline{K}^{*0}}$-parameter, could be accommodated simultaneously.  All our results and the small parameter space could be further tested in the running LHC experiments, Belle-II and future high-energy colliders.
\end{abstract}
\newpage
\section{Introduction} \label{sec:1}
It is well known that $B$ meson rare decays provide us an abundant source of information on QCD, $CP$ violation and new physics (NP) beyond the Standard Model (SM). In recent years, the anomalies such as $R(D^{(*)})$ and $R_{K^{(*)}}$ observed in semileptonic $B$ meson rare decays at large hadron collider (LHC) and $B$-Factories imply that the lepton flavour universality may be violated, which in particular are viewed as the signals of the effects of NP (for recent reviews, see, e.g., Refs.~\cite{Li:2018lxi, Bifani:2018zmi, London:2021lfn, Altmannshofer:2021qrr}). Unlike the semileptonic decays, the hadronic $B$ decays suffer from larger uncertainties and are therefore more difficult to calculate with a high accuracy, because the hadronic matrix elements cannot be calculated from the first principle directly. In the past twenty years, based on the factorization hypothesis \cite{Bauer:1986bm}, some QCD based approaches to handle such kinds of problems are usually discussed in the heavy quark limit and implemented by the heavy quark expansion, such as the light-cone sum rule (LCSR) \cite{Khodjamirian:2000mi}, the QCD factorization (QCDF) \cite{Beneke:1999br, Beneke:2003zv}, the soft-collinear effective theory (SCET) \cite{Bauer:2000yr, Bauer:2001cu} and the perturbative QCD (PQCD) factorization approach \cite{Keum:2000ph, Lu:2000em, Ali:2007ff}. However, the observables such as the branching fractions, $CP$ asymmetries, polarization fractions and angular distributions might suffer from large uncertainties from higher-order and higher-power contributions. In this sense, in hadronic $B$ decays a deviation with respect to the SM prediction requires one to be much more conservative regarding these uncertainties than in the case of semileptonic $B$ decays. For this reason, in order to search for the signals of NP in the hadronic heavy flavour particle decays, on the one hand we should reduce the theoretical uncertainties as possible by preforming the higher order and higher power corrections with the developments of QCD technique, but on the other we are encouraged to search for new observables that are insensitive to the theoretical uncertainties.

Among the two-body $B$ meson hadronic decays, it is of great interest to us that the decays $B_{d} \to K^{*0} {\overline K^{*0}}$ and $B_{s} \to K^{*0} {\overline K^{*0}}$ have same final states and are related by $U$-spin. Both two decays are induced by the flavor-changing neutral-current (FCNC) transitions, in which new particles of NP could affect the observables by entering the loops. In addition, $B_{s} \to K^{*0} {\overline K^{*0}}$ decay is also regarded as a golden channel for a precision measurement of the CKM phase $\beta_s$ \cite{Ciuchini:2007hx}. In the experimental side, both the branching fractions and the longitudinal polarization fractions have been measured in two $B$ factories \cite{Aubert:2007xc, BaBar:2007wwj, Belle:2010uya} and LHCb experiment \cite{LHCb:2011btn, LHCb:2015exo, Aaij:2017wgt, Aaij:2019loz}. For the decay $B_{d} \to K^{*0} {\overline K^{*0}}$, the theoretical predictions of the branching fraction and polarization fractions based on QCDF \cite{Beneke:2006hg} and PQCD \cite{Zou:2015iwa,Chai:2022kmq} are all in agreement with the averaged experimental results \cite{Workman:2022ynf}  $B(B_{d} \to K^{*0} {\overline K^{*0}})=(8.3 \pm 2.4) \times 10^{-7}$ and $f_L(B_{d} \to K^{*0}  {\overline K^{*0}}) = 0.74 \pm 0.05 $ within the large uncertainties. Furthermore, the measurement of $f_L(B_{d} \to K^{*0} {\overline K^{*0}})$ agrees with the na\"ive hypothesis, based on the quark helicity conservation and the $(V-A)$ nature of the weak interaction. For the decay $B_{s} \to K^{*0} {\overline K^{*0}}$, the latest averaged experimental results \cite{Workman:2022ynf} are $B(B_{s} \to K^{*0} {\overline K^{*0}}) = (11.1\pm 2.7) \times 10^{-6}$, $f_L(B_{s} \to K^{*0}  {\overline K^{*0}}) = 0.240 \pm 0.031 \pm0.025 $ and $f_\perp(B_{s} \to K^{*0} {\overline K^{*0}}) = 0.38 \pm 0.11 \pm 0.04$. It is found that the prediction of branching fraction $B(B_{s} \to K^{*0} {\overline K^{*0}}) = (9.1^{+0.5+11.3}_{-0.4-6.8}) \times 10^{-6}$ in QCDF \cite{Beneke:2006hg} agrees well with the data, but the longitudinal polarization fraction $f_L(B_{s} \to K^{*0}  {\overline K^{*0}}) =0.63^{+0.42}_{-0.29}$ is much larger than the data. On the another side, based on PQCD approach \cite{Zou:2015iwa}, the predicted branching fraction and longitudinal polarization fraction are $B(B_{s} \to K^{*0} {\overline K^{*0}})=(5.4^{+3.0}_{-2.4}) \times 10^{-6}$  and $f_L(B_{s} \to K^{*0}{\overline K^{*0}}) =0.38^{+0.12}_{-0.10}$, respectively. It is seen that although the longitudinal polarization fraction $f_L$ is consistent with the data, its center value is a bit smaller than the experimental measurement. Altogether, the theoretical predictions with large uncertainties from two approaches cannot explained all available data convincingly. In order to explain the current data simultaneously, the theoretical predictions with high precision in both approaches are called, and we are also encouraged to explore the contributions of NP.

Following \cite{Descotes-Genon:2011rgs}, the authors in ref.\cite{Alguero:2020xca} defined an observable that is sensitive to the $U$-spin asymmetry but with a cleaner theoretical prediction as
\begin{equation}\label{eq:LK}
L_{K^{*0}\overline{K}^{*0}}=\frac{{B}(B_{s}\to K^{*0}\overline{K}^{*0}) g(B_{s}\to K^{*0}\overline{K}^{*0}) f_L(B_{s}\to K^{*0}\overline{K}^{*0})}{{B}(B_{d}\to K^{*0}\overline{K}^{*0}) g(B_{d}\to K^{*0}\overline{K}^{*0}) f_L(B_{d}\to K^{*0}\overline{K}^{*0})},
\end{equation}
where the phase space factors $g(B_{Q}\to K^{*0}\overline{K}^{*0})$ involved in the corresponding branching fractions are given as
\begin{equation}
g(B_{Q}\to K^{*0}\overline{K}^{*0})= \frac{\tau_{B_Q}}{16\pi M_{B_Q}^2} \sqrt{M_{B_Q}^2-4M_{K^{*0}}^2}\,.
\end{equation}
In such a ratio, the experimental uncertainties are reduced, as the uncertainties in the denominator and numerator can be cancelled out by each other. In \cite{Aaij:2019loz}, LHCb collaboration released the measurements of the ratio between two branching fractions and the longitudinal polarization fraction of $B_{s} \to K^{*0} {\overline K^{*0}}$. With the latest results and the longitudinal polarization fraction of $B_{d} \to K^{*0} {\overline K^{*0}}$ from PDG \cite{Workman:2022ynf}, we could obtain this new observable as
\begin{equation}\label{eq:expL}
L_{K^*\overline{K}^*}^{\rm Exp}=4.43\pm 0.92,
\end{equation}
where the effect of $B_s$ meson mixing in the measurement of the branching fraction is included. In QCDF, the prediction based on the results from \cite{Beneke:2006hg} is given as \cite{Alguero:2020xca}
\begin{eqnarray}\label{eq:tension1}
 L_{K^*\overline{K}^*}^{\rm QCDF}=& 19.5^{+9.3}_{-6.8},
  \label{eq:tension3}
  \end{eqnarray}
which implies a $2.6\sigma$ tension with respect to the experimental data. This new ``anomaly" discrepancy is viewed as a new signal of NP \cite{Alguero:2020xca}. However, $L_{K^*\overline{K}^*}$ of PQCD is not available yet till now. Motivated by this, we  shall exploit this observable in PQCD in this work and  try to check whether the $L_{K^*\bar{K}^*}$ is still lager than the experimental data. Moreover, the branching fractions and polarizations of both two decays will also be recalculated with the new fitted distribution amplitudes of $K^*$ \cite{Hua:2020usv}.

As aforementioned, in order to interpret the called $R_K$ and $R_{K^*}$ anomalies, large number of NP models have been proposed. One of the most popular NP explanations are models with an extra heavy vector $Z^\prime$ boson \cite{Albrecht:2021tul, Geng:2021nhg}, where the new introduced $Z^\prime$ boson has couplings to quarks, as well as to either electrons or muons with non-universal parameters. In order to test these models, besides searching $Z^\prime$ at the higher energy colliders directly, the signals in other observables involving the similar transitions are also expected. A straightforward place to explore the possible existence of these signals are hadronic $B$ decays induced by the FCNC transitions $b\to (d,s) q\bar q$. In SM, such kind of decays are forbidden at tree level and only occur by loops. The comparable contributions from $Z^\prime$ at tree level may change the observables remarkably. Hence, another purpose of this work is to explore whether the contributions of an extra $Z^\prime$ boson can explain all measured observables in some certain spaces of parameters.

This paper is organized as follows. We will first present the calculations of $B_{d} \to K^{*0} {\overline K^{*0}}$ and $B_{s} \to K^{*0} {\overline K^{*0}}$ decays in SM within the PQCD approach, and more attentions are mainly paid on not only branching fractions and the longitudinal polarization fractions but the new observable $ L_{K^*\overline{K}^*}$. In Sec.\ref{sec:3}, we will study contributions from the non-universal $Z^\prime$ boson, which could change the observables in the suitable parameters space. Lastly, we shall summarize this work in Sec. \ref{sec:4}.

\section{Calculation in SM} \label{sec:2}
In SM, the decay amplitudes of of $B_{d,s}\to K^{*0} {\overline K^{*0}}$ decays follow from the matrix elements $\langle V_{2}V_{3}|H_\text{eff}| B\rangle$ of the effective Hamiltonian
\begin{equation}  \label{Hamiltonian}
H_\text{eff} =
     \frac{G_F}{\sqrt{2}} \sum_{p=u,c} \lambda^{(D)}_p
     \left\{
        C_{1} Q_{1}^p + C_{2} Q_{2}^p +\!\!
        \sum_{i=3,\ldots 10}\!\! C_i Q^p_i
     \right\} + \mathrm{h.\,c},
\end{equation}
with $D \in \{d,s\}$ and $\lambda^{(D)}_p = V_{pb}^*V_{pD}$. $C_{i}(\mu)$ are Wilson coefficients, and $O_{i}(\mu)(i=1,2,3 \cdots, 10)$ are the four-quark effective operators, whose specific forms refer to \cite{Buchalla:1995vs}.

In PQCD, the $B$ meson amplitude  can be expressed as \cite{Keum:2000ph}
\begin{eqnarray} \label{PQCD}
\langle V_{2}V_{3}\left|{H}_\text{eff}\right|B\rangle
&\sim& \int  dx_{1}dx_{2}dx_{3}b_{1}db_{1}b_{2}db_{2}b_{3}db_{3}\nonumber\\
&&\times{\rm Tr}\left[C(t)\Phi_{B}(x_{1},b_{1})\Phi_{V_{2}}(x_{2},b_{2})\Phi_{V_{3}}(x_{3},b_{3})H(x_{i},b_{i},t)S_{t}(x_{i})e^{-S(t)}\right].
\end{eqnarray}
The meson wave functions $\Phi_i$ ($i=B,V_2,V_3$) include the dynamical information that how the quarks are combined into a hadron. They are nonperturbative but universal. $\rm Tr$ is the sum of degrees of freedom in the spin and color space. $b_{i}$ is the conjugate variable of the quark transverse momentum $k_{iT}$, and $x_{i}$ is the longitudinal momentum fraction carried by the light quark in each meson. $H(x_{i},b_{i},t)$ describes the four quark operators and the spectator quark connected by a hard gluon, and can be calculated perturbatively. The jet function $S_{t}(x_{i})$ coming from the threshold resummation of the double logarithms $\ln^2 x_i$ smears the end-point singularities in $x_{i}$ \cite{Li:2001ay}. The Sudakov form factor  $e^{-S(t)}$ arising from the resummation of the double logarithms suppresses the soft dynamics effectively i.e. the long distance contributions in the large-$b$ region \cite{Li:1994iu, Keum:2000wi}. The main advantage of this approach is that it preserves the transverse momenta of quarks and avoids the problem of end-point divergence.

Because there are three kinds of polarizations for a vector meson, namely longitudinal ($L$), normal ($N$) and transverse ($T$),  the amplitudes for a $B$ meson decay to two vector mesons are generally characterized by the polarization states of two vector mesons. Thus, the amplitude $ A^{(\sigma)}$ for the decay $B(P_B) \to V_2(P_2,\epsilon_{2\mu}^{*}) V_3(P_3,\epsilon_{3\mu}^{*})$ can be decomposed as follows:
\begin{eqnarray}
A^{(\sigma)}& =&\epsilon_{2 \mu}^{*}(\sigma)\epsilon_{3\nu}^{*}(\sigma)\left[a g^{\mu \nu}+\frac{b}{M_{2} M_{3}} P_{B}^{\mu} P_{B}^{v}+i \frac{c}{M_{2} M_{3}} \epsilon^{\mu \nu \alpha \beta} P_{2 \alpha} P_{3 \beta}\right]\nonumber\\
&=&A_{L}+A_{N} \epsilon_{2}^{*}(\sigma=T) \cdot \epsilon_{3}^{*}(\sigma=T)+i\frac{A_{T}}{M_{B}^{2}} \epsilon^{\mu \nu \gamma \rho} \epsilon_{2 \mu}^{*}(\sigma) \epsilon_{3 v}^{*}(\sigma) P_{2 \gamma} P_{3 \rho},
\end{eqnarray}
where $M_2$ and $M_3$ are the masses of the vector mesons $V_2$ and $V_3$, respectively. The definitions of the amplitudes $A_{i}$ $(i=L,N,T)$ in terms of the Lorentz-invariant amplitudes $a$, $b$ and $c$ could be written as
\begin{align}
A_{L}&=a \epsilon_{2}^{*}(L) \cdot \epsilon_{3}^{*}(L)+\frac{b}{M_{2}M_{3}} \epsilon_{2}^{*}(L) \cdot P_{3} \epsilon_{3}^{*}(L) \cdot P_{2},\\
A_{N}&=a,\\
A_{T}&=\frac{c}{r_{2} r_{3}},
\end{align}
with $r_{2,3}=M_{V_{2,3}}/M_B$. The amplitudes $ A_{i}$ $(i=L,N,T)$ could be calculated in PQCD approach directly.

Alternatively, we can also define the polarization amplitudes of three directions, and their relationships with $A_{L}$, $A_{N}$ and $A_{T}$ are given as follows:
\begin{equation}
A_{0}=-A_{L}, \,\,\,A_{\|}=\sqrt{2}A_{N},\,\,\, A_{\perp}=r_{2}r_{3} \sqrt{2\left(\kappa^{2}-1\right)} A_{T},
\end{equation}
with the ratio $\kappa=\frac{P_{2} \cdot P_{3}}{M_{K^{*0}}}$. Then, the branching fraction of $B\to V_2V_3$ is expressed as
\begin{align} \label{br}
{B}\left(B\to V V\right)=\tau_{B} \frac{\left|p_{c}\right|}{8 \pi M_{B}^{2}}\left[\left|A_{0}\right|^{2}+\left|A_{\|}\right|^{2}+\left|A_{\perp}\right|^{2}\right],
\end{align}
where $\tau_B$ is the lifetime of the $B$ meson, $p_c$ is the three-dimension momentum of the vector meson. Three polarization fractions $f_i(i=L, \parallel, \perp)$ are also defined as
\begin{eqnarray}\label{pvf}
f_i=\frac{|A_i|^2}{|A_0|^2+|A_\parallel|^2+|A_\perp|^2}\;.
\end{eqnarray}


In PQCD approach, the most important inputs are the wave functions of hadrons. For the initial state $B$ meson, its wave function is of the form \cite{Zou:2015iwa,Ali:2007ff,Xiao:2006hd,Li:2004ep}
\begin{equation}
 \Phi_{B}(x,b) = \frac{i}{\sqrt{2N_c}}
\left[ \not \! P_{B} \gamma_5 + M_{B} \gamma_5 \right]
\phi_{B}(x,b),
\end{equation}
where $b$ is the conjugate space coordinate of the transverse momentum $k_\perp$, and $N_c=3$ is the number of color. The distribution amplitude $\phi_{B}$ is in the form of
\begin{align}
\phi_{B}(x, b)=N_{B}x^{2}(1-x)^{2} \exp \left[-\frac{1}{2}\left(\frac{x m_{B}}{\omega_{B}}\right)^{2}-\frac{\omega_{B}^{2} b^{2}}{2}\right],
\end{align}
where $N_{B}$ is the normalization factor and satisfies \begin{align}
\int_{0}^{1} d x \phi_{B}(x, b=0)=\frac{f_{B}}{2 \sqrt{2 N_{c}}},
\end{align}
$f_{B}$ being the decay constant of $B$ meson. The shape parameter $\omega_{B}=0.30$ and $\omega_{B_s}=0.50$ are determined by experimental data or calculated from the first principle \cite{Wang:2019msf}.

Unlike the pseudoscalar particle, the vector meson has the longitudinal polarization vector $\epsilon_L$ and the transverse polarization one $\epsilon_T$. For a special final state $K^{*0}$ moving in the plus direction ($n_{+}$) with momentum $P$, two wave functions of the $K^{*0}$ up to twist-3 are given as \cite{Ball:2007rt}
\begin{eqnarray}
\Phi_{K^*}^\parallel &=&\frac{1}{\sqrt{2N_c}}\left[M_{K^*}\not\!\epsilon_{L}\phi_{K^*}(x)+\not\!\epsilon_{L}\not\!P\phi_{K^*}^t(x)+M_{K^*}\phi_{K^*}^s(x)\right],\\
\Phi_{K^*}^\perp &=&\frac{1}{\sqrt{2N_c}}\left[ M_{K^*}\not\! \epsilon^*_T\phi_{K^*}^v(x)+ \not\!\epsilon^*_T\not\!P\phi_{K^*}^T(x)+iM_{K^*}\epsilon_{\mu\nu\rho\sigma}\gamma_5\gamma^\mu\epsilon_T^{*\nu}n_+^\rho n_-^\sigma \phi_{K^*}^a(x)\right ],
\end{eqnarray}
where $n_{+}=\left(1,0, \mathbf{0}_{T}\right)$ and $n_{-}=\left(0,1, \mathbf{0}_{T}\right)$. Two polarizations are defined as
\begin{align}
\epsilon(L)=\frac{P}{M_{K^{*}}}-\frac{M_{K^{*}}}{P \cdot n_{+}} n_{+},\,\,\,\,\,\,
\epsilon(T)=\left(0,0,\mathbf{1}_{T}\right),
\end{align}
The light-cone distribution amplitudes in the wave function have been calculated within the QCD sum rules \cite{Ball:2005vx, Ball:2007zt},
\begin{eqnarray}
\phi_{K^{*}}(x)&=&\frac{3f_{K^{*}}}{\sqrt{2N_c}} x(1-x)\left[1+a_{1K^{*}}^{\|} C_{1}^{3/2}(t)+a_{2K^{*}}^{\|} C_{2}^{3 / 2}(t)\right],
\\
\phi_{K^{*}}^{T}(x)&=&\frac{3f_{K^{*}}}{\sqrt{2N_c}} x(1-x)\left[1+a_{1K^{*}}^{\perp} C_{1}^{3/2}(t)+a_{2 K^{*}}^{\perp} C_{2}^{3/2}(t)\right],
\\
\phi_{K^{*}}^{t}(x)&=&\frac{3f_{K^{*}}^{T}}{2\sqrt{2N_c}}t^{2},
\\
\phi_{K^{*}}^{s}(x)&=&\frac{3f_{K^{*}}^{T}}{2\sqrt{2N_c}}(-t),
\\
\phi_{K^{*}}^{v}(x)&=&\frac{3f_{K^{*}}}{8\sqrt{2N_c}}\left(1+t^{2}\right),\\
\phi_{K^{*}}^{a}(x)&=&\frac{3 f_{K^{*}}}{4\sqrt{2N_c}}(-t).
\end{eqnarray}
The Gegenbauer polynomials in the distribution amplitude are given as
\begin{align}
C_{1}^{3/2}(t)=3 t,\quad C_{2}^{3/2}(t)=\frac{3}{2}\left(5 t^{2}-1\right),
\end{align}
where $t=2x-1$ and $x$ is the momentum fraction of the light quark.

\begin{figure}
\begin{center}
\includegraphics[scale=0.75]{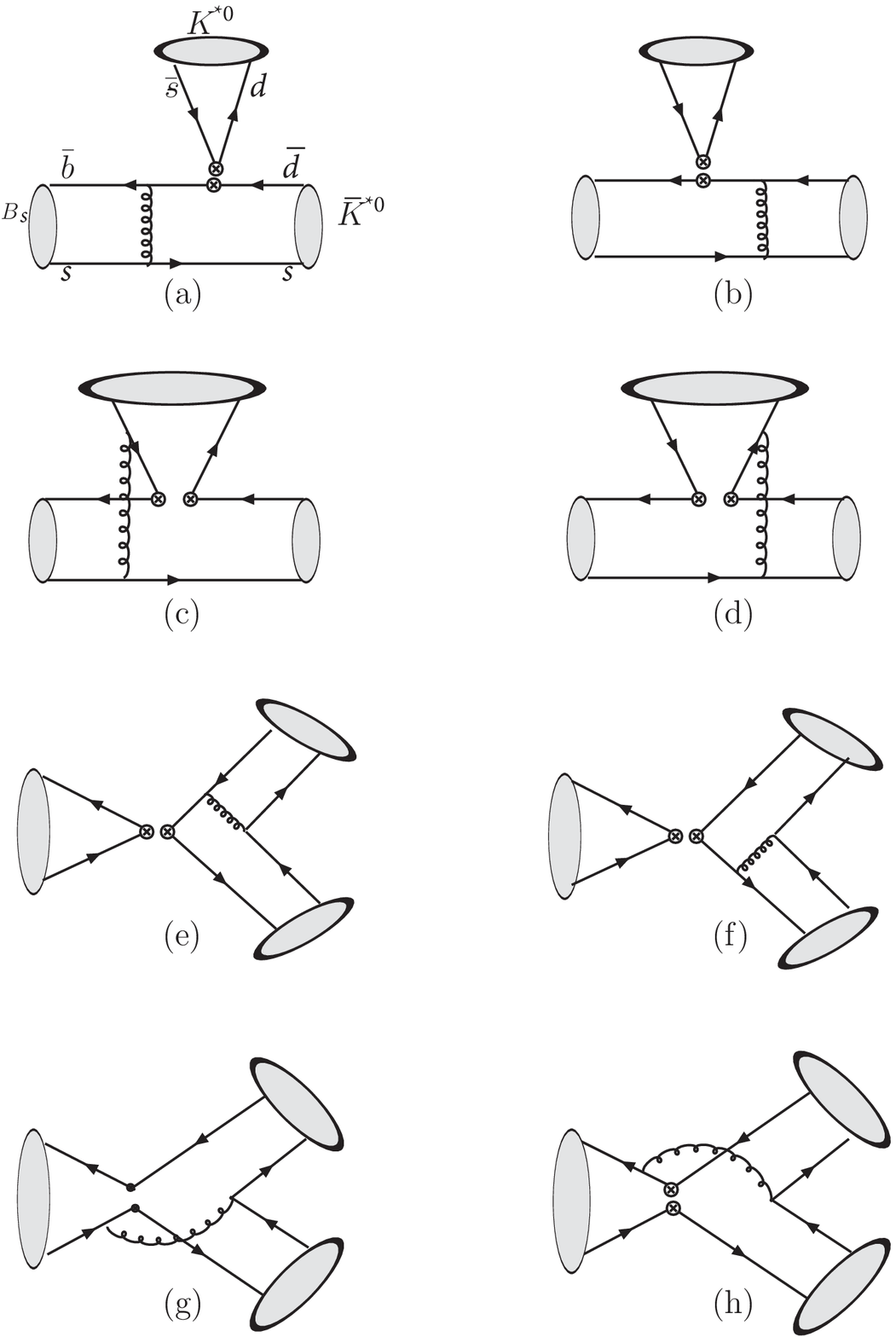}
\caption{The leading order Feynman diagrams for $B_{s} \to K^{*0} {\overline K^{*0}}$}\label{Feynman Diagram}
\end{center}
\end{figure}

According to the effective Hamiltonian eq.(\ref{Hamiltonian}), we could draw the lowest order diagrams contributing to $B_{d,s}\to K^{*0} \overline{K}^{*0}$. For example, the Feynman diagrams of $B_s\to K^{*0} \overline{K}^{*0}$ are shown in Fig.~\ref{Feynman Diagram}, where the symbols ``$\otimes$" are the effective operators. The figures (a) and (b) are factorizable emission diagrams, while (c) and (d) are nonfactorizable emission ones. Similarly, figures (e) and (f) are factorizable annihilation diagrams, and (g) and (h) are nonfactorizable annihilation ones. We also note that in $B_s\to K^{*0} \overline{K}^{*0}$ decay the final vector meson $\overline{K}^{*0}$ takes the spectator strange quark, while in $B_d\to K^{*0} \overline{K}^{*0}$ decay the spectator down quark enters ${K}^{*0}$ meson.

After calculating the amplitudes of each diagram with different operators, we obtain the amplitudes of $B^0 \to K^{*0} {\overline K^{*0}}$ and  $B^0_{s} \to K^{*0} {\overline K^{*0}}$, which are given as
\begin{align}
A^{i}\left(B^{0} \rightarrow K^{*0} \overline{K}^{*0}\right)=&-\frac{G_{F}}{\sqrt{2}} V_{tb}^{*} V_{t d}\left\{ M_{fh}^{LL, i}\left[a_{4}-\frac{1}{2} a_{10}\right]+M_{nfh}^{LL,i}\left[C_{3}-\frac{1}{2} C_{9}\right]\right.+M_{nfh}^{LR,i}\left[C_{5}-\frac{1}{2}C_{7}\right]
\nonumber\\
&
 +M_{fa}^{LL,i}\left[\frac{4}{3}a_{3}+\frac{4}{3}a_{4}-\frac{2}{3}a_{9}-\frac{2}{3} a_{10}\right] +M_{fa}^{LR,i}\left[a_{5}-\frac{1}{2} a_{7}\right]+M_{fa}^{SP,i}\left[a_{6}-\frac{1}{2} a_{8}\right] \nonumber\nonumber\\
&+M_{nfa}^{LL,i}\left[C_{3}-\frac{1}{2}C_{9}+C_{4}
 -\frac{1}{2}C_{10}\right]  +M_{nfa}^{LR,i}\left[C_{5}-\frac{1}{2}C_{7}\right]
 +M_{nfa}^{SP,i}\left[C_{6}-\frac{1}{2} C_{8}\right] \nonumber\\
&+\left(M_{fa}^{LL,i}\left[a_{3}-\frac{1}{2} a_{9}\right]+M_{fa}^{LR, i}\left[a_{5}-\frac{1}{2} a_{7}\right]\right)_{K^{*0} \leftrightarrow \overline{K}^{*0}} \nonumber\\
&\left.+\left(M_{nfa}^{LL,i}\left[C_{4}-\frac{1}{2} C_{10}\right]+M_{nfa}^{SP,i}\left[C_{6}-\frac{1}{2}C_{8}\right]\right)_{K^{*0} \leftrightarrow \overline{K}^{*0}}\right\},\label{BDKK}
\end{align}
\begin{align}
A^{i}\left(B_{s}^{0}\rightarrow K^{*0}\overline{K}^{*0}\right)=&-\frac{G_{F}}{\sqrt{2}}V_{tb}^{*} V_{ts}\left\{ M_{fh}^{LL, i}\left[a_{4}-\frac{1}{2} a_{10}\right]+M_{nfh}^{LL,i}\left[C_{3}-\frac{1}{2} C_{9}\right]+M_{nfh}^{LR,i}\left[C_{5}-\frac{1}{2}C_{7}\right]\right.
\nonumber\\
& +M_{fa}^{LL,i}\left[\frac{4}{3}a_{3}+\frac{4}{3}a_{4}-\frac{2}{3}a_{9}-\frac{2}{3} a_{10}\right]+M_{fa}^{LR,i}\left[a_{5}-\frac{1}{2} a_{7}\right]+M_{fa}^{SP,i}\left[a_{6}-\frac{1}{2} a_{8}\right] \nonumber\nonumber\\
&+M_{nfa}^{LL,i}\left[C_{3}-\frac{1}{2}C_{9}+C_{4}
 -\frac{1}{2}C_{10}\right] +M_{nfa}^{LR,i}\left[C_{5}-\frac{1}{2}C_{7}\right]
 +M_{nfa}^{SP,i}\left[C_{6}-\frac{1}{2} C_{8}\right] \nonumber\\
&+\left(M_{fa}^{LL,i}\left[a_{3}-\frac{1}{2} a_{9}\right]+M_{fa}^{LR, i}\left[a_{5}-\frac{1}{2} a_{7}\right]\right)_{K^{*0} \leftrightarrow \overline{K}^{*0}} \nonumber\\
&\left.+\left(M_{nfa}^{LL,i}\left[C_{4}-\frac{1}{2} C_{10}\right]+M_{nfa}^{SP,i}\left[C_{6}-\frac{1}{2}C_{8}\right]\right)_{K^{*0} \leftrightarrow \overline{K}^{*0}}\right\},\label{BSKK}
\end{align}
with
\begin{align}
a_{1}=C_{2}+C_{1}/3,\quad a_{2}&=C_{1}+C_{2}/3,\quad a_{3}=C_{3}+C_{4}/3,\quad a_{4}=C_{4}+C_{3}/3, \nonumber\\
a_{5}=C_{5}+C_{6}/3,\quad a_{6}&=C_{6}+C_{5}/3,\quad a_{7}=C_{7}+C_{8}/3,\quad a_{8}=C_{8}+C_{7}/3, \nonumber\\
a_{9}&=C_{9}+C_{10}/3,\quad a_{10}=C_{10}+C_{9}/3,
\end{align}
where $i=L,N,T$ denote the longitudinal polarization and the two transverse polarizations. In above two formulae, the superscripts $LL$, $LR$ and $SP$ indicate the operators $(V-A)(V-A)$, $(V-A)(V+A)$ and $(S-P)(S+P)$, respectively. The subscript ``$fh$" in $M_{fh}$ meas factorizable emission diagrams $(a)$ and $(b)$, while ``$nfh$" means nonfactorizable ones $(c)$ and $(d)$. Similarly, ``$fa$" and ``$nfa$" are the the factorizable and nonfactorizable annihilation diagrams, respectively. Due to the limit of space, we will not list the above amplitudes for each $M$, and the explicit expressions can be found in refs.~\cite{Ali:2007ff, Zou:2015iwa}. It should be stressed that all amplitudes ``$M$" are mode dependent, as the spectator quarks are different in these two decays, though the eqs.(\ref{BDKK}) and (\ref{BSKK})  are very similar.

With above formulae, we then calculate the observables in SM.  The branching fractions and longitudinal polarization fractions of both decays are given in Table.~\ref{tab:1}, together with predictions of QCDF and the available experimental data. In our numerical calculations, the updated distribution amplitudes \cite{Hua:2020usv} of $K^*$ are adopted. We acknowledge that there are still some uncertainties in our calculations, and we here only discuss two main uncertainties. In the table, the first errors arise from the wave functions of heavy $B$ mesons, in which the shape parameters $ \omega_{B_d}$ and $\omega_{B_s}$ are the only inputs, and we make them change $30\%$. The second ones are from the next-leading power (order) corrections characterized by the hard scale $t$, which changes from $0.8 t$ to $1.2 t$. It can be seen that the branching fractions are affected by both parameters, while the polarization fractions are only sensitive to the shape parameter $\omega_{B_d}$ or $\omega_{B_s}$. In PQCD, both $B_d^0 \to K^{*0}\overline{K}^{*0}$ and $B_{s}^0 \to K^{*0} \overline{K}^{*0}$ are induced only by the penguin operators, so that the direct $CP$ asymmetries of two decays are zero in PQCD. However, including the contributions from charm penguins, the direct $CP$ asymmetries from QCDF are nonzero. Thus, the measurements of direct $CP$ asymmetries in future could discriminate two approaches.

\begin{table}[!htp]
\begin{center}
    \caption{Numerical results for observables in $B_{d,s} \to K^{*0} {\overline K^{*0}}$ decays in SM, tegather with results of QCDF and experimental results.}\label{tab:1}
    \begin{tabular}{ccccc}
    \hline\hline
     Decay Mode
     & BF $(10^{-6})$
     &$f_{L}(\%)$
     &$f_{\|}(\%)$
     &$f_{\perp}(\%$)\\ \hline

     $B^{0} \rightarrow K^{*0}\overline{K}^{*0}$
     &$0.5_{-0.1-0.1}^{+0.2+0.2}$
     &$67.1_{-5.7-0.4}^{+5.1+0.3} $
     &$17.4_{-3.4-0.0}^{+3.6+0.1}$
     &$15.5_{-2.5-0.2}^{+2.7+0.1}$ \\  \hline
     QCDF \cite{Beneke:2006hg}
     &$0.6_{-0.1-0.3}^{+0.1+0.5}$
     &$69_{-1-27}^{+1+34}$ \\  \hline
     Exp. \cite{Workman:2022ynf}
     &$0.8 \pm 0.09 \pm 0.04$
     &$72.4 \pm 5.1\pm 1.6 $
     &$11.6 \pm 3.3\pm 1.2$
     &$16\pm 4.4\pm 1.2 $ \\
     \hline
     \hline
     $B_{s}^{0} \rightarrow K^{* 0} \overline{K}^{* 0}$
     &$7.8_{-1.4-1.5}^{+1.9+2.3}$
     &$51.1_{-6.8-0.3}^{+7.3+0.6} $
     &$25.6_{-4.2-0.3}^{+3.7+0.1} $
     &$23.3_{-3.5-0.2}^{+3.3+0.3} $ \\  \hline
     QCDF \cite{Beneke:2006hg}
     &$9.1^{+0.5+11.3}_{-0.4-6.8}$
     &$63_{-0-29}^{+0+42}$ \\  \hline
     Exp. \cite{Workman:2022ynf}
     &$11.1\pm 2.2\pm1.2$
     &$24\pm 3.1\pm 2.5$
     &$ $
     &$38_{-11-4}^{+11+4}$
     \\
    \hline\hline
    \end{tabular}
    \end{center}
\end{table}

From Table.~\ref{tab:1}, we find that for the decay $B^{0} \rightarrow K^{*0}\overline{K}^{*0}$, the predictions of branching fractions and polarization fractions from PQCD and QCDF are in agreement with the experimental results, though the theoretical center values of branching fraction are smaller than the experimental data. In fact, the longitudinal contribution is dominant, which is roughly proportional to the form factor $A_0^{B\to K^*}$. In QCDF, $A_0^{B\to K^*}(0)=0.39\pm 0.06$ calculated from light-cone sum rules \cite{Ball:2004rg} was adopted, while $A_0^{B\to K^*}(0)=0.36\pm 0.05$ is obtained in PQCD. In addition, the form factors $A_1^{B\to K^*}(0)$ and $V^{B\to K^*}(0)$ that are relate to transverse amplitudes are almost same in PQCD and QCDF. For the decay $B_s^{0} \to K^{*0}\overline{K}^{*0}$, the theoretical predictions are in agreement with each other with uncertainties, with $A_0^{B_s\to K^*}(0)=0.33\pm 0.05$ in QCDF and $A_0^{B_s\to K^*}(0)=0.30\pm 0.05$ in PQCD. However, in comparison to the experimental results, both branching fractions are smaller than the data, and both theoretical longitudinal polarization fractions are much larger than data, even the predictions of QCDF have large uncertainties arising from annihilation diagrams. In our previous study \cite{Zou:2015iwa}, with the large suppression from threshold resummation, the predicted longitudinal polarization fraction $f_L=(38.3^{+12.1}_{-10.5})\%$ could be comparable to data, but the corresponding branching fraction $(5.4^{+3.0}_{-2.4})\times 10^{-6}$ is smaller than the current data. Although there are many uncertainties in the theoretical calculations, this discrepancy could be a hint of NP beyond SM.

Now, we calculate the $L_{K^*\overline{K}^{*0}}$-parameter and  obtain
\begin{eqnarray}
 L_{K^*\overline{K}^{*0}}^{\rm PQCD}= 12.7^{+5.6}_{-3.2},\label{LKKPQCD}
  \end{eqnarray}
where the uncertainty is mainly from the shape parameters in the distribution amplitudes of $B^0_d$ and $B^0_s$ mesons. The uncertainties taken by high order corrections are almost cancelled. In this sense, the more precise and reliable shape parameters of heavy mesons based on the nonperturbative approaches are needed. By comparison, we find our result is also larger than one from the current data, eq.(\ref{eq:expL}), though it is smaller than that of QCDF.

\section{Calculation in Family Nonuniversal $Z^{\prime}$ Model} \label{sec:3}
Now, we turn to study the contributions of the extra gauge boson $Z^\prime$ to the decays $B^0_{s} \to K^{*0} {\overline K^{*0}}$ which is induced by the FCNC $b\to s \bar d d$ transition. Supposing there is no mixing between $Z$ and $Z^\prime$, the $Z^\prime$ term of the neutral-current Lagrangian in the gauge basis can be written as \cite{Langacker:2000ju,Langacker:2008yv}
\begin{eqnarray}
L^{Z^\prime} =-g^\prime Z^{\prime {\mu}}\sum_{i} {\overline \psi_i^I} \gamma_{\mu} \left[ (\epsilon_{\psi_L})_{i} P_L + (\epsilon_{\psi_R})_{i} P_R \right] \psi^I_j,
\end{eqnarray}
where $\psi^I_i$ means the $i$-th family fermion, and the superscript $I$ refers to the gauge interaction eigenstate. $g^\prime$ is the gauge coupling constant at the electro-weak scale $M_W$, and $P_{L,R}=(1\mp\gamma_5)/2$. The parameter $\epsilon_{\psi_L}$ ($\epsilon_{\psi_R}$) denotes the left-handed (right-handed) chiral coupling. According to certain string constructions \cite{Chaudhuri:1994cd} or GUT models \cite{Barger:1987hh}, the couplings can be family non-universal. When we change the weak basis to the physical one, FCNC's generally appear at tree level in both left-handed and right-handed sectors, explicitly, as
\begin{eqnarray}
B^{L}=V_{\psi_L}\epsilon_{\psi_L}V_{\psi_L}^{\dagger},\;\;\;\;\;
B^{R}=V_{\psi_R}\epsilon_{\psi_R}V_{\psi_R}^{\dagger},
\end{eqnarray}
where $V_{\psi_{L,R}}$ are unitary matrices. For simplicity, the right-handed couplings are supposed to be flavor-diagonal. Therefore, the FCNC $b\to s\bar{q}q$ (and $q=u,d$) transition can also be mediated by the $Z^{\prime}$ at tree level, and the corresponding effective Hamiltonian has the form as:
\begin{equation}\label{heffz1}
 {H}_{eff}^{\rm Z^{\prime}}=\frac{2G_F}{\sqrt{2}}\big(\frac{g^{\prime}M_Z}
 {g_1M_{Z^{\prime}}}\big)^2\,B_{sb}^L(\bar{s}b)_{V-A}\sum_{q}\big(B_{qq}^L (\bar{q}q)_{V-A} +B_{qq}^R(\bar{q}q)_{V+A}\big)+h.c.\,,
\end{equation}
where $g_1=e/(\sin{\theta_W}\cos{\theta_W})$ and $M_{Z^{\prime}}$ is the mass of the new $Z^\prime$ boson. The current structures $(V-A)(V-A)$ and $(V-A)(V+A)$, are same as eq.(\ref{Hamiltonian}) of SM, which allow us to translate eq.~(\ref{heffz1}) as
\begin{equation}
 {H}_{eff}^{\rm
 Z^{\prime}}=-\frac{G_F}{\sqrt{2}}V_{tb}V_{ts}^{\ast}\sum_{q}
 (\Delta C_3 O_3^q +\Delta C_5 O_5^q+\Delta C_7 O_7^q+\Delta C_9
  O_9^q)+h.c.\,.
\end{equation}
In above Hamiltonian, $\Delta C_i$ denote $Z^{\prime}$ corrections to the Wilson coefficients of the SM operators, which can be written as
\begin{eqnarray}
 \Delta C_{3}&=&-\frac{2}{3V_{tb}V_{ts}^{\ast}}\,\big(\frac{g^{\prime}M_Z}
 {g_1M_{Z^{\prime}}}\big)^2\,B_{sb}^L\,(B_{uu}^{L}+2B_{dd}^{L})\,,\nonumber\\
 \Delta C_{5}&=&-\frac{2}{3V_{tb}V_{ts}^{\ast}}\,\big(\frac{g^{\prime}M_Z}
 {g_1M_{Z^{\prime}}}\big)^2\,B_{sb}^L\,(B_{uu}^{R}+2B_{dd}^{R})\,,\nonumber\\
 \Delta C_{7}&=&-\frac{4}{3V_{tb}V_{ts}^{\ast}}\,\big(\frac{g^{\prime}M_Z}
 {g_1M_{Z^{\prime}}}\big)^2\,B_{sb}^L\,(B_{uu}^{R}-B_{dd}^{R})\,,\nonumber\\
 \Delta C_{9}&=&-\frac{4}{3V_{tb}V_{ts}^{\ast}}\,\big(\frac{g^{\prime}M_Z}
 {g_1M_{Z^{\prime}}}\big)^2\,B_{sb}^L\,(B_{uu}^{L}-B_{dd}^{L})\,.
 \label{NPWilson}
\end{eqnarray}
It is obvious that $Z^\prime$ contributes to the QCD penguins as well as to the EW penguins. For simplicity, we follow the assumptions in refs.~\cite{Buras:2003dj, Barger:2009eq, Hua:2010wf, Chang:2013hba, Li:2015xna, Chang:2009wt, Celis:2015ara} and set $B_{uu}^{L,R}=-2 B_{dd}^{L,R}$, so that new physics is manifest in the EW penguins, namely $O_7$ and $O_9$. Furthermore, without loss of generality, the diagonal elements of the effective coupling matrices $B_{qq}^{L,R}$ are supposed to be real due to the hermiticity of the effective Hamiltonian. However, there is no constrain that the off-diagonal $B_{sb}^{L}$ should be a real, and a new weak phase $\phi_{bs}$ can exist. Taking all these information together, we then have the new  Wilson coefficients
\begin{eqnarray}
& &\Delta C_{3,5}\simeq 0, \nonumber\\
& &\Delta
C_{9,7}=4\frac{|V_{tb}V_{ts}^{\ast}|}{V_{tb}V_{ts}^{\ast}}\xi^{L,R}e^{i\phi_{bs}},
\end{eqnarray}
with
\begin{eqnarray}\label{xi}
\xi^{L,R}=\left(\frac{g^{\prime}M_Z}
 {g_1M_{Z^{\prime}}}\right)^2\left|\frac{B_{sb}^LB_{dd}^{L,R}}{V_{tb}V_{ts}^{\ast}}
 \right|.
\end{eqnarray}

With the assumption that both ${U(1)_{Y}}$ in the SM and ${U(1)}$ introduced in new models origin from the Grand Unified Theory, the gauge coupling constants for ${Z}$ and ${Z^{\prime}}$ bosons are the same, implying that $g^\prime/g_1=1$. So far, the obvious signal of the new ${Z^{\prime}}$ boson have not been observed in the current experiments such as CMS and ATLAS, which indicates that the mass of $Z^\prime$ would be larger than the Tev scale. Conservatively, we set $M_{Z} / M_{Z^{\prime}} \approx 0.1$. In order to accommodate the mass difference between $B_{s}^0$ and $\overline{B}_{s}^0$ that is one of the  most strictest constraints to the models with $Z^\prime$ boson, $\left|B_{sb}^{L}\right| \sim\left|V_{tb} V_{ts}^{*}\right|$ is theoretically required. Meanwhile, in order to explain $CP$ asymmetries of $B \to K \pi$ and branching fractions of $B \to K \phi$ and $B \to K^* \phi$, the diagonal elements should satisfy $\left|B_{qq}^{L,R}\right|\sim 1$. For the newly introduced weak phase $\phi_{bs}$, it is assumed to be a free parameter without any restriction whose range is $[-\pi, \pi]$. In order to reduce the number of new parameters, we further assume $\xi=\xi^{L L}=\xi^{L R}$, which means that the left-hand couplings are same as right-handed ones. Of course, $\xi^{L L}=0$ or $\xi^{L R}=0$ can  also assumed, and we shall not discuss these two cases any more. Therefore, in our following discussion, we have only two parameters $\xi \in[0.001,0.02]$ and $\phi_{bs} \in [-180^\circ,180^\circ]$.

\begin{figure}[htb]
\begin{center}
\includegraphics[scale=0.8]{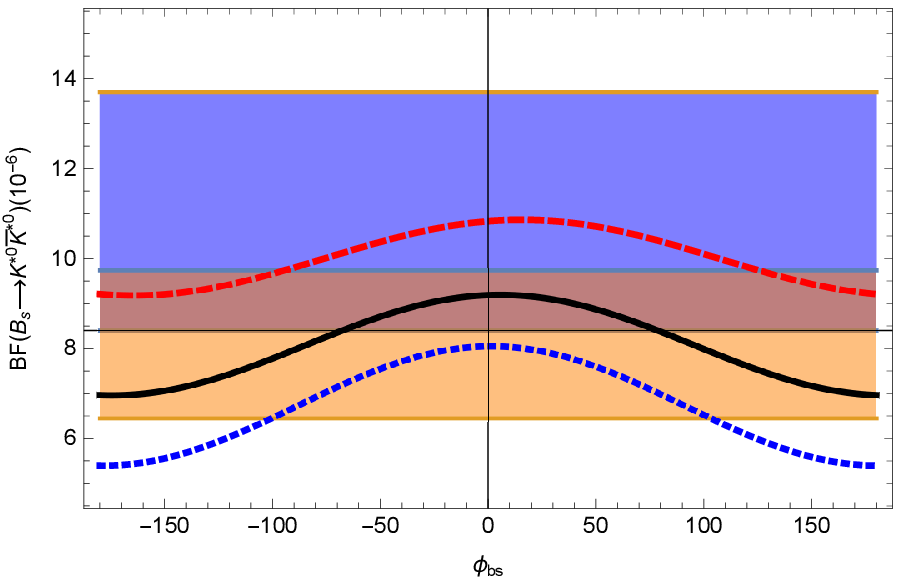}
\includegraphics[scale=0.8]{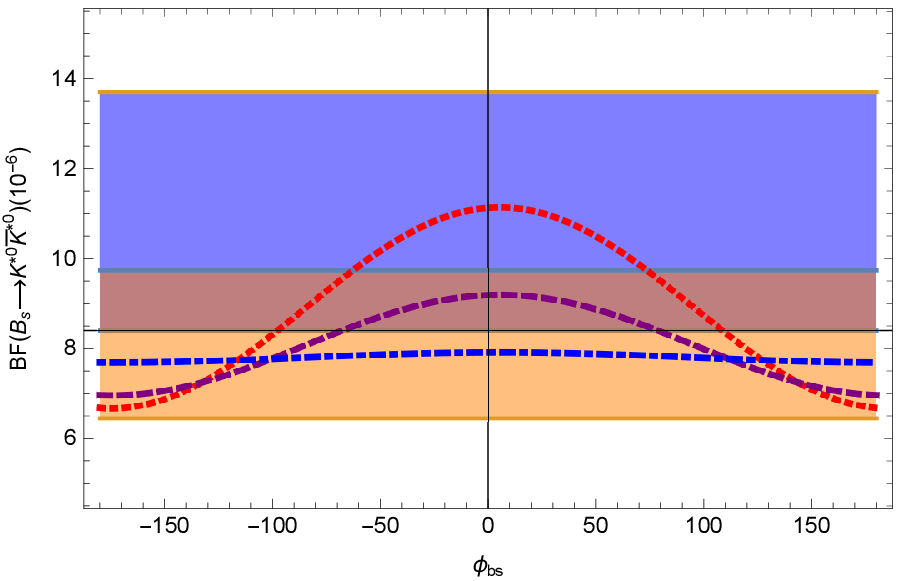}
\caption{The dependence of the branching fraction of $B_s\to K^{*0}\overline{K}^{*0}$ on the weak phase $\phi_{bs}$, for a fixed value $\xi=0.01$ with $\omega_{B_s}=0.45$ (dotted blue line), $0.50$ (solid black line) and $0.55$ (dashed red line) in the left panels; and for a fixed $\omega_{B_s}=0.50$ with $\xi= 0.005$ (dot-dashed blue line) , $0.01$ (dashed purple line)  and $0.02$ (dotted red line) in the right panels. The blue and yellow regions represent the experimental data and SM prediction, respectively.}\label{fig:2}
\end{center}
\end{figure}

\begin{figure}[htb]
\begin{center}
\includegraphics[scale=0.8]{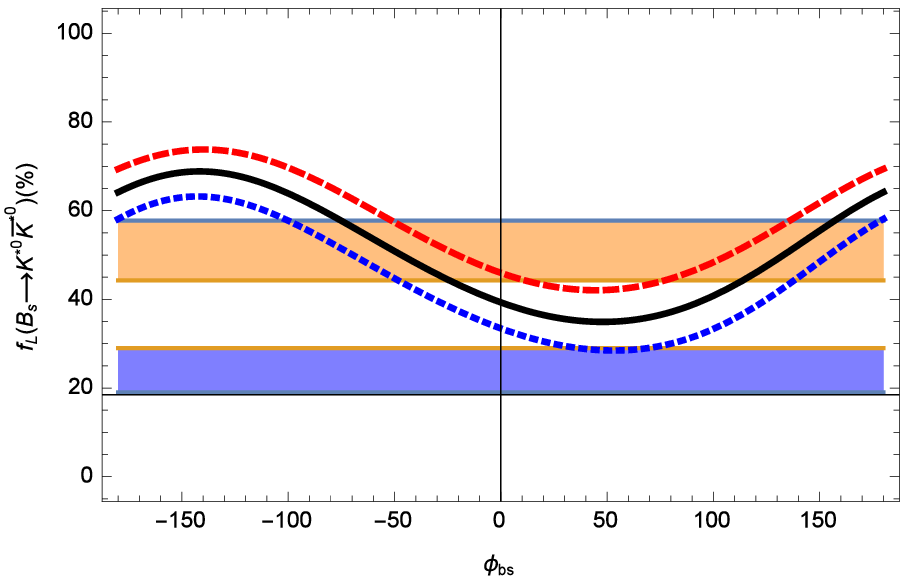}
\includegraphics[scale=0.8]{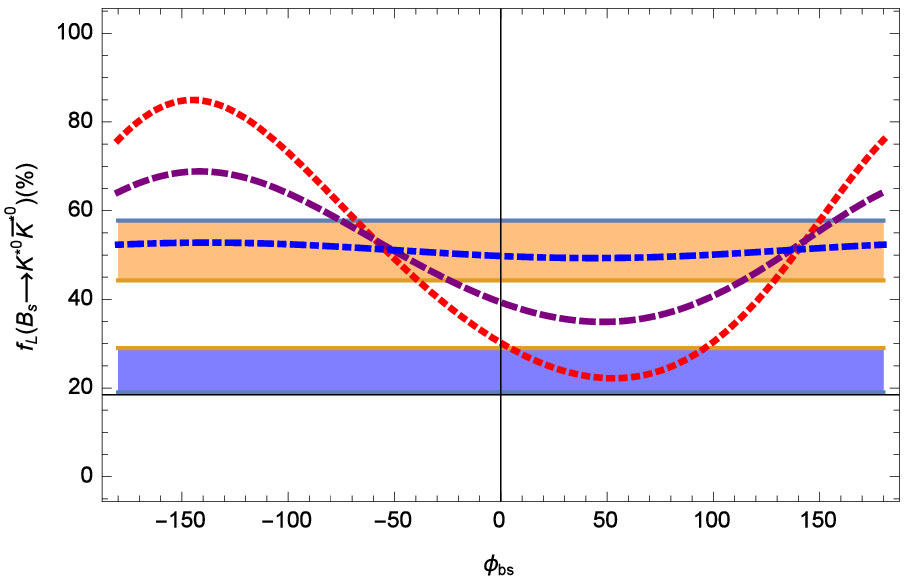}
\caption{The dependence of the longitudinal polarization fraction ($f_L$) of $B_s\to K^{*0}\overline{K}^{*0}$ on the weak phase $\phi_{bs}$, for a fixed value $\xi=0.01$ with $\omega_{B_s}=0.45$ (dotted blue line), $0.50$ (solid black line) and $0.55$ (dashed red line) in the left panel; and for a fixed $\omega_{B_s}=0.50$ with $\xi= 0.005$ (dot-dashed blue line) , $0.01$ (dashed purple line)  and $0.02$ (dotted red line) in the right panel. The blue and yellow regions represent the experimental data and SM prediction, respectively. }\label{fig:3}
\end{center}
\end{figure}

In Figure.~\ref{fig:2} and Figure.~\ref{fig:3}, we present the branching fraction and longitudinal polarization fraction of $B_s\to K^{*0}\overline{K}^{*0}$ as functions of the new weak phase $\phi_{bs}$, for a fixed value $\xi=0.01$ with $\omega_{B_s}=0.45, 0.50$ and $0.55$ in the left panels, and for a fixed $\omega_{B_s}=0.50$ with $\xi=0.02,0.01$ and $0.005$ in the right panels. The experimental data and the SM predictions are also shown in the figures for comparisons. As aforementioned, the experimental result and theoretical prediction of SM on the branching fraction have some overlaps, but there is no overlap on the longitudinal polarization fraction. From Table.~\ref{tab:1}, we could see that in SM the uncertainty of the branching fraction arising from the $\omega_{B_s}$ is about $20\%$. With the fixed parameter $\xi=0.01$, for each $\omega_{B_s}$, the uncertainties coming from the unknown phase $\phi_{bs}$ are also around $20\%$, as shown in the left panel of Figure.~\ref{fig:2}. Comparing the theoretical results with the data, a small $\omega_{B_s}$ is preferred by the experimental data. Given $\omega_{B_s}=0.50$, it is found from the right panel of Figure.~\ref{fig:2} that if $\xi<0.01$ the contributions of the new particle would be plagued by the large theoretical uncertainties. However, when we set $\xi=0.02$, the effect from $Z^\prime$ boson becomes more remarkable, and the branching fraction could be as large as $11.2\times 10^{-6}$ when $\phi_{bs}=0^{\circ}$. Specifically, for $\xi=0.02$ and $\omega_{B_s}=0.50$, the new weak phase $\phi_{bs}$ is constrained in the range $[-100^{\circ},100^{\circ}]$ by the current data, and the range decreases as $\xi$ becomes smaller.

In contrast to the branching fraction, the measured longitudinal polarization fraction is smaller than the theoretical prediction, which allows us to find out some mechanisms to suppress the longitudinal contribution or enhance the transverse contributions. It can be seen from the left panel of Fig.~\ref{fig:3} that for the fixed value $\xi=0.01$, most results are larger than the data, and only few results approach the upper limit of experimental data when $\omega_{B_s}=0.55$ and $\phi_{bs}\approx50 ^\circ$. Therefore, a larger $\omega_{B_s}$ is favored, which is different from the result from the well measured branching fraction. It is shown in the right panel that, for the fixed $\omega_{B_s}=0.50$, the theoretical predictions of longitudinal polarization fractions $f_L$ are larger than the data, for both $\xi=0.01$ and $\xi=0.001$. When $\xi=0.02$, $f_L$ changes in a wide range with the changes of $\phi_{bs}$, and could fall into the experimental range within $\phi_{bs}\in [8 ^\circ,93 ^\circ]$. When $\phi_{bs}\approx50 ^\circ$, $f_L$ could be as small as $22\%$.

\begin{figure}[!ht]
\begin{center}
\includegraphics[scale=0.8]{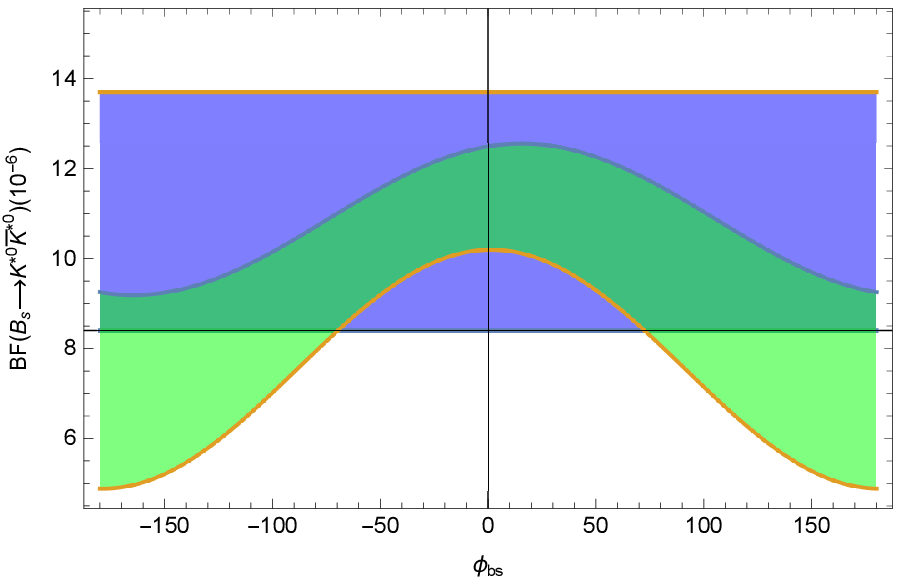}\hspace{0.4cm}
\includegraphics[scale=0.8]{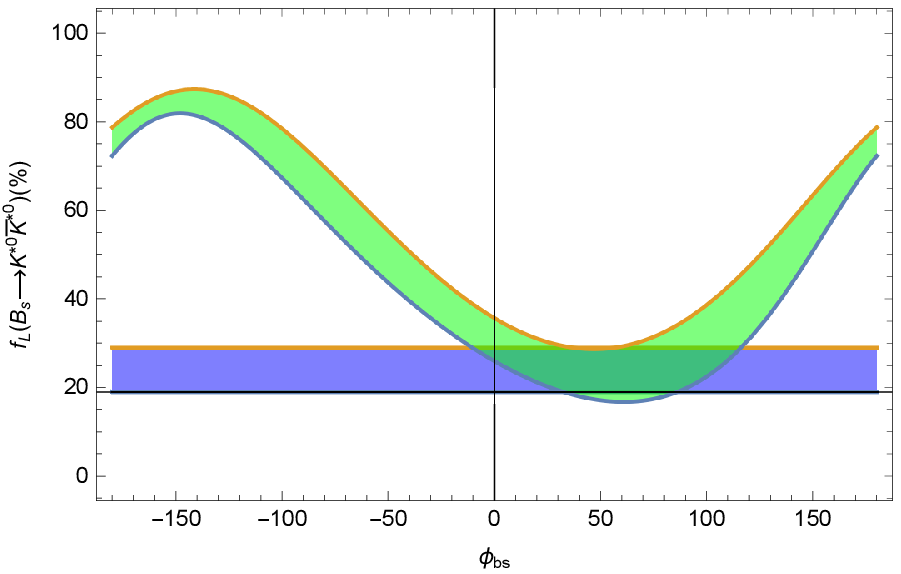}
\caption{The dependence of the branching fraction (left panel) and longitudinal polarization fraction ($f_L$) (right panel) of $B_s\to K^{*0}\overline{K}^{*0}$ on the weak phase $\phi_{bs}$, for a fixed value $\xi=0.02$ with $\omega_{B_s}=0.50\pm0.05$. The blue bands represent the experimental data. }\label{fig:4}
\end{center}
\end{figure}

From above analysis, the branching fraction prefers a smaller $\omega_{B_s}$, while the longitudinal polarization fraction prefers a larger one. Also, we found that once $\xi=0.02$ is adopted, both the branching fraction and the longitudinal polarization fraction vary in a large region with the change of $\phi_{bs}$. Thus, with $\xi=0.02$ we plot all possible regions for $\omega_{B_s}=0.50\pm0.05$ in Fig.~\ref{fig:4}. These two figures illustrate that for the fixed $\xi=0.02$ both two observables could be consistent with the experimental data well, even $\omega_{B_s}=0.45$ is adopted. In addition, a positive weak phase $\phi_{bs}$ is preferred, as implied in Fig.~\ref{fig:4}.

\begin{figure}[htb]
\begin{center}
\includegraphics[scale=0.8]{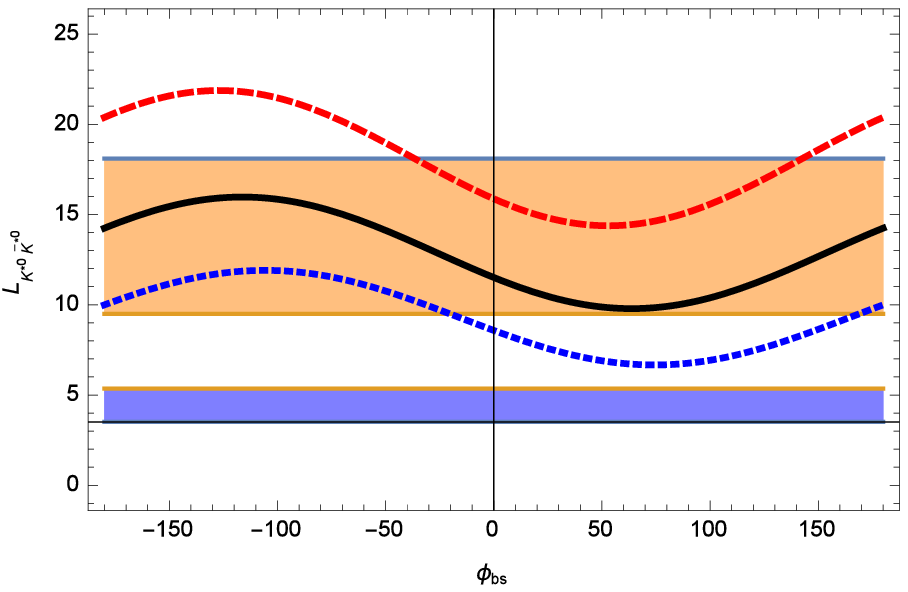}\hspace{0.4cm}
\includegraphics[scale=0.8]{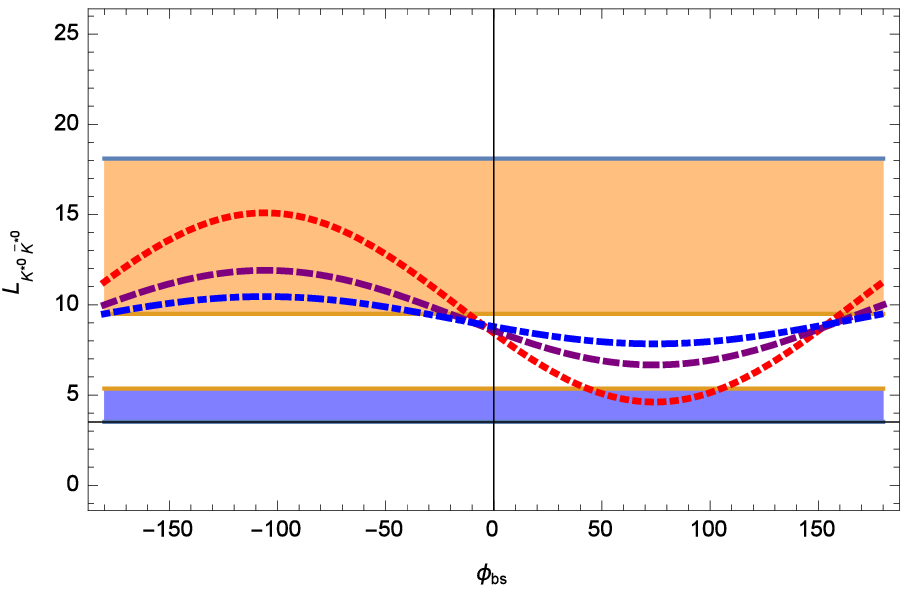}
\caption{The dependence of $L_{K^*\overline{K}^{*0}}$-parameter  on the weak phase $\phi_{bs}$, for a fixed value $\xi=0.01$ with $\omega_{B_s}=0.45$ (dotted blue line), $0.50$ (solid black line) and $0.55$ (dashed red line) in the left panel, and for a fixed $\omega_{B_s}=0.50$ with $\xi= 0.005$ (dot-dashed blue line) , $0.01$ (dashed purple line)  and $0.02$ (dotted red line) in the right panel. The blue and yellow regions represent the experimental data and SM prediction, respectively.}\label{fig:5}
\end{center}
\end{figure}

\begin{figure}[htb]
\begin{center}
\includegraphics[scale=0.8]{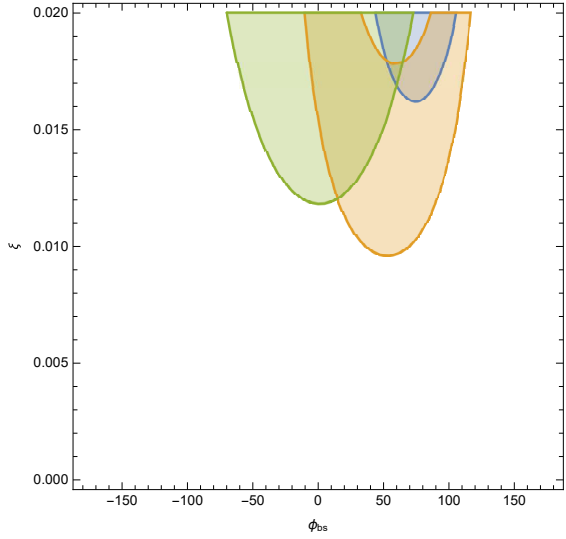}
\caption{Combined constraints on the $(\phi_{bs},\xi)$ two-dimensional plane for the fixed value $\omega_{B_s}=0.55$. The green, yellow and blue regions represent the constraints from the branching fraction and  the longitudinal polarization fraction of $B_s\to K^{*0}\overline{K}^{*0}$ decay,  and $L_{K^*\overline{K}^{*0}}$-parameter, respectively.}\label{fig:6}
\end{center}
\end{figure}

Now, we shall discuss the effect of the new introduced $Z^\prime$ boson on the new defined parameter $L_{K^*\overline{K}^{*0}}$. As aforementioned, we suppose that $Z^\prime$ only participates in the $b\to s$ transitions, and its contribution to the FCNC $b\to d$ transitions is suppressed by small $|B_{db}|$ and negligible. In this respect, $L_{K^*\overline{K}^{*0}}$ does in fact reflect the contribution of longitudinal amplitude of decay $B_{s}^{0} \to K^{*0} {\overline{K}^{*0}}$. In the left panel of Figure.~\ref{fig:5}, we adopt $\xi=0.01$ again and show the variant of $L_{K^*\overline{K}^{*0}}$ with changes of $\phi_{bs}$ for $\omega_{B_s}=0.45, 0.50$ and $0.55$. The SM prediction and the latest measurement are also shown. By comparison, we find that if $\xi=0.01$ the theoretical predictions cannot agree with  experimental data, even if $\omega_{B_s}=0.55$ is adopted. By setting $\omega_{B_s}=0.45, 0.50$ and $0.55$, we also calculated  $L_{K^*\overline{K}^{*0}}$. The numerical results show that if $\xi<0.02$ the values of $\omega_{B_s}=0.45, 0.50$ are not preferred by the experimental data. Thus, we adopt $\omega_{B_s}=0.55$ and plot $L_{K^*\overline{K}^{*0}}$ dependence on the phase for $\xi=0.001, 0.01$ and $0.02$ in the right panel. It can be clearly seen that $L_{K^*\overline{K}^{*0}}$ changes in a wide range  for $\xi=0.02$, and it could be 4.61 as $\phi_{bs} \approx 75^\circ$. Combining Figures.~\ref{fig:4} and ~\ref{fig:5}, we find that in such a family non-universal $Z^\prime$ model there might exist a certain parameter space, where all observables can be achieved. In order to obtain the parameter space, we show the combined result in the $(\phi_{bs},\xi)$ two-dimensional plane for the fixed value $\omega_{B_s}=0.55$, as shown Figure~\ref{fig:6}. The green and yellow bands represent the regions fitting the branching fraction and the longitudinal polarization fraction respectively,  while the region of the parameter space corresponding to a viable fit of $L_{K^*\overline{K}^{*0}}$ has been marked in blue. Evidently, the experimental data of $L_{K^*\overline{K}^{*0}}$ gives the most stringent constraint. As was expected, these three bands overlap in a very small region,  $\xi \in [0.017,0.018]$ and $\phi_{bs}\in [50^\circ,65^\circ]$. Within this small parameter space, we then have
\begin{eqnarray}
B(B_{s}^{0} \to K^{* 0} \overline{K}^{* 0})&=&(8.6\pm0.4)\times 10^{-6}, \\
f_{L}(B_{s}^{0} \to K^{* 0} \overline{K}^{* 0})&=&(19.5\pm0.7)\%\\
L_{K^*\overline{K}^{*0}}^{\rm PQCD}&=&5.3\pm 0.3.
\end{eqnarray}
These results with few uncertainties could be further tested with high precision in the current LHCb experiment or the Belle-II experiment.

Finally, we present some comments on the direct searches of $Z^\prime$ boson. At the LHC, the  main way to search directly for a $Z^\prime$ is via a resonance peak in the invariant-mass distribution of its decay products. This experimental analysis is usually performed by the ATLAS and CMS collaborations for $Z^\prime$ production in the $s$-channel in a rather model-independent way, but assuming that the observed new resonance is narrow, such that any interference of SM and NP contributions can be neglected. Under these assumptions, the $Z^\prime$ Drell-Yan cross section at a hadron machine can be approximated as \cite{Accomando:2010fz,Paz:2017tkr, Workman:2022ynf}
\begin{eqnarray}
\sigma(pp\to Z^\prime X \to f\bar{f} X) \simeq \frac{\pi}{6 s} \sum_q c_q^f w_q(s,{M_{Z^\prime}}^2)
\end{eqnarray}
where $q=u,d,s,c,b$. Here, the hadronic structure functions $w_q(s,{M_{Z^\prime}}^2)$ are independent of the $Z^\prime$ model and contain all information on parton distribution functions and QCD corrections. On the other hand, the coefficients $c_q^f$ contain all model-dependent information. Recently, ATLAS and CMS collaborations published the limits on $M_{Z^\prime}$ as a function of $c_{u,d}^\ell$ where $\ell =e,\mu$ \cite{ATLAS:2019erb, CMS:2021ctt}. The lower mass limits of $5.15 (4.56)$ TeV are set based on the sequential standard model (superstring-inspired model) \cite{CMS:2021ctt}, and the lower limits could reach $4.5$ TeV for the $E_6$-motivated $Z^\prime$ boson \cite{ATLAS:2019erb}. However, our results are challenged by above measurements, because the combined parameter $\xi \in [0.017,0.018]$ implies that the large $g^\prime$ or small $M_{Z^\prime}$ are needed, as shown in eq.(\ref{xi}). We also note for high values of $g^\prime$ the ratio $g^\prime/M_{Z^\prime}$ can be quite large, which could spoil the narrow-width approximation. Besides, the current limits are all model-dependence, and the model-independent analyses are not available yet. Therefore the models with $M_{Z^\prime}\leq 3-4$ TeV required by flavour physics cannot be excluded totally by current data. We look forward to further searches of $Z^\prime$ in the current LHC experiment  or future high-energy colliders.
\section{Summary}\label{sec:4}
In this work, we studied the nonleptonic decays $B_{d} \to K^{*0} {\overline K^{*0}}$ and $B_{s} \to K^{*0} {\overline K^{*0}}$ within the perturbative QCD approach, which is based on the $k_{\rm T}$ factorization. With the new fitted distribution amplitudes of $K^{*}$, both the branching fractions and the polarization fractions are recalculated. Numerical results show that the theoretical results of $B_{d} \to K^{*0} {\overline K^{*0}}$ are in agreement with experimental measurements, while for the decay $B_{s} \to K^{*0} {\overline K^{*0}}$ the branching fraction and the longitudinal polarization fraction cannot agree with data simultaneously. We also explored the $L_{K^*\overline{K}^{*0}}$-parameter that is a combination of polarization fractions and branching fractions in order to reduce the theoretical uncertainties. In SM, $L_{K^*\overline{K}^{*0}}^{\rm PQCD}= 12.7^{+5.6}_{-3.2}$ is obtained based on PQCD, which is still larger than the experimental data. In order to identify whether the deviations come from the contribution of new physics, the accuracy of theoretical calculations should be further improved in future, for example  exploring the wave function of heavy $B$ meson. On the other side, we are also encouraged to search for the effects of NP beyond SM. Then, we interpreted these deviations by introducing a family nonuniversal $Z^{\prime}$ boson in $b\to s q\bar q$ transition. In order to reduce the number of new parameters, we simplified the model as possible. With the large shape parameter $\omega_{B_s}=0.55$ in the distribution amplitude of $B_s$ meson, it is in a small parameter space $\xi \in [0.017,0.018]$ and $\phi_{bs}\in [50^\circ,65^\circ]$ that these three measurements (branching fraction, longitudinal polarization fraction and $L_{K^*\overline{K}^{*0}}$-parameter) could be accommodated simultaneously. In such small parameter space, the theoretical uncertainties could be reduced remarkably. All our results are hopeful tested in LHCb experiment, Belle-II and future high-energy colliders.
\section*{Acknowledgment}
This work is supported in part by the National Science Foundation of China under the Grant Nos. 11975195, 11365018 and 11375240, and the Natural Science Foundation of Shandong province under the Grant No.ZR2019JQ04. This work is also supported by the Project of Shandong Province Higher Educational Science and Technology Program under Grants No.2019KJJ007.
{\small
\bibliographystyle{bibstyle}
\bibliography{mybibfile}
}
\end{document}